\documentclass[aip,apl,twocolumn,preprintnumbers,amsmath,amssymb]{revtex4}

\usepackage{graphicx}
\usepackage[colorlinks=true,citecolor=blue]{hyperref}

\newcommand{\unit}[1]{\hspace{0pt} {#1}}

\def\ket#1{\mathinner{|{#1}\rangle}}
\def\braket#1{\mathinner{\langle{#1}\rangle}}

\begin{document}
    \title{Ultrafast photon-photon interaction in a strongly coupled quantum dot-cavity system}
    \author{Dirk Englund$^{1,2,\dag}$}
    \email{englund@columbia.edu}
    \author{Arka Majumdar$^{1,\dag}$}
    \email{arkam@stanford.edu}
    \author{Michal Bajcsy$^1$}
    \author{Andrei Faraon$^1$}
    \author{Pierre Petroff$^3$}
    \author{Jelena Vu\v{c}kovi\'{c}$^1$}
    \affiliation{$^1$E.L.Ginzton Laboratory, Stanford University, Stanford, CA, 94305\\}
 \affiliation{$^2$Department of Electrical Engineering and Department of Applied Physics, Columbia University, New York, NY 10027\\}
    \affiliation{$^3$Materials Department, University of California, Santa Barbara, CA
    93106\\}
    \affiliation{$^\dag$ These authors contributed equally\\}

\begin{abstract}
We study dynamics of the interaction between two weak light beams
mediated by a strongly coupled quantum dot-photonic crystal cavity
system. First, we perform all optical switching of  a weak
continuous-wave signal with a pulsed control beam, and then
perform switching between two pulsed beams (40ps pulses) at the
single photon level. Our results show that the quantum
dot-nanocavity system creates strong, controllable interactions at
the single photon level.
\end{abstract}
\maketitle Techniques to efficiently interact single photons with
quantum emitters are fundamental to the field of quantum optics
and quantum information and are at the core of a range of
protocols, including two-qubit phase
gates\cite{1995.PRL.Turchette-Kimble,1995.PRA.Chuang-Yamamoto.simple_QC}
and quantum non-demolition measurements\cite{1985.PRA.Yamamoto}.
In addition, single-photon-level optical nonlinearities may enable
ultra-low power and high-speed all-optical gates and switches for
classical optical information processing
\cite{1989.book.Tooley-Wherrett.optical_comp,2009.PRA.Mabuchi.cavity_QED_switch}.
Recent experiments have shown that the necessary nonlinearity may
be realized using atomic gases in the slow light
regime\cite{2009.PRL.Bajcsy-Vuletic.fiber_switching} or
solid-state systems consisting of a single quantum dot (QD)
strongly coupled to a nano-cavity
\cite{2007.Nature1,2007.Nature.Srinivasan-Painter,2008.Science1}.
As a crucial next step in the solid state approach, we describe
here the time-resolved dynamics of light interacting the QD-cavity
system. We first study the  the interaction of a weak
continuous-wave signal and a pulsed control, and then the
interaction of two short ($40$ ps) pulses.

\begin{figure}[]
\centering{\includegraphics[width=3.5in]{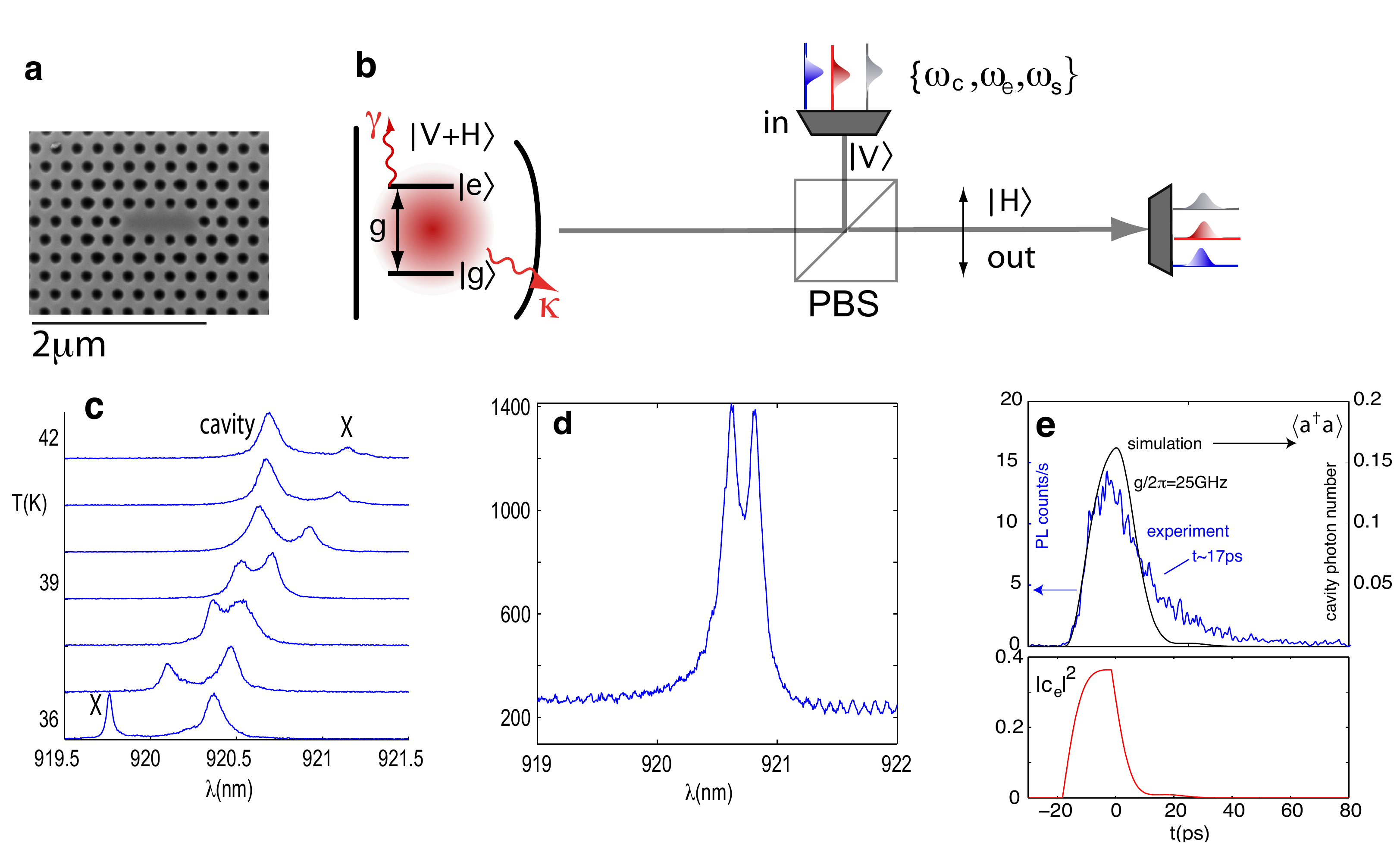}}
\caption{{\footnotesize (a) Scanning electron micrograph of the
photonic crystal cavity. (b) In the experiment, a combination of
pulsed control laser (frequency $\omega_c$), non-resonant pulsed
or continuous wave pump laser at $\omega_e$ above the QD exciton
line, and pulsed or continuous-wave signal laser ($\omega_s$) is
employed. The cavity is backed by a distributed Bragg reflector,
effectively creating a single-sided cavity\cite{2007.Nature1}. A
cross-polarized confocal microscope configuration reduces the
laser background that is reflected from the sample without
coupling to the cavity, which is polarized at $45^{\circ}$ to the
incident laser polarization. The measurement is effectively a
transmission measurement from the horizontal (H) into the vertical
(V) polarization.  (c) Anticrossing observed in the PL as the QD
single exciton (X) is temperature-tuned through the cavity. The QD
is pumped through higher-order excited states by optical
excitation at $\omega_e$ corresponding to a wavelength of
$\lambda_e=878$ nm. (d) The reflected intensity of a broad-band
light source shows the mode splitting of the strongly coupled
QD-cavity system. The reflected light is resolved on a
spectrometer. (e) PL lifetime $\sim 17$\unit{ps} when the QD is
tuned into the cavity and pulsed excitation wavelength
$\lambda_e=878$\unit{nm}. The emission that is expected
theoretically, based on the system parameters in (d) and a 10-ps
relaxation time into the single exciton state, is shown in the
solid line. The bottom panel plots the corresponding expected
excited state population $|c_e(t)|^{2}$} }\label{fig:setup}
\end{figure}

The experiment is performed using a QD dot strongly coupled to
three-hole (L3) photonic crystal (PC) cavity
\cite{Noda2003Nature}, superposed with a grating to increase the
emission directionality \cite{2009.OpEx.Toishi-Englund} (see
Fig.\ref{fig:setup}(a)). It was fabricated in a $160$ nm thick
membrane containing a central layer of self-assembled InAs QDs
with a density of $\sim 50/\mu\mbox{m}^{2}$ and an inhomogeneously
distributed exciton emission between 925$\pm 15$ \unit{nm}.

The eigen-frequencies  $\omega_{\pm}$ of the QD-cavity system are:
\begin{equation}
\omega_{\pm}=\frac{\omega_r+\omega_d}{2}-i\frac{\kappa+\gamma}{2}
\pm \sqrt{g^2+\frac{1}{4}\left(\delta-i(\kappa-\gamma)\right)^2}
\label{eq:eigenstates}
\end{equation}
where $\omega_r$ and $\omega_d$ are the cavity and QD resonance
frequencies, respectively; $\kappa$ and $\gamma$ are the cavity
field decay rate and QD dipole decay rate; $g$ denotes the
coherent interaction strength between the QD; and
$\delta=\omega_d-\omega_r$ is the cavity-QD detuning.  The
parameters of the emitter-cavity system used in the experiment are
$g/2\pi=25$ GHz, $\kappa/2\pi=27$ GHz, $\gamma/2\pi = 0.1 $ GHz.
Therefore, the expression under square root in
Eq.~\ref{eq:eigenstates} is positive for $\delta=0$, implying that
the system is in the strong coupling regime of the cavity quantum
electrodynamics (QED).

We characterize the system in a confocal microscope setup in a He
flow cryostat (Fig.\ref{fig:setup}(b)). The photoluminescence (PL)
scans in Fig.\ref{fig:setup}(c) show the anticrossing between the
QD-like states and the cavity-like states as the temperature is
raised from 36K to 42 K, giving the polariton energies given by
Eq.\ref{eq:eigenstates}. The cavity reflectivity, obtained using a
white light source in the cross-polarized configuration shown in
Fig.\ref{fig:setup}(b), shows the same mode splitting in
Fig.\ref{fig:setup}(d). These spectral measurements yield the
system parameters $g,\kappa$, and $\gamma$. We also characterize
the system by time-resolved fluorescence after the QD is
quasi-resonantly excited with 3.5-ps pulses at 878\unit{nm}. As
shown in Fig.\ref{fig:setup}(e), the PL decays with a
characteristic time of 17\unit{ps}, as measured using a streak
camera with 3 ps timing resolution. This decay closely matches a
theoretical model of the cavity field and coupled QD system, as
described in the Methods. The bottom panel of
Fig.\ref{fig:setup}(e) plots the expected excited state population
$|c_e(t)|^{{2}}$, which shows a rise-time corresponding to the
10-ps carrier relaxation time into the lowest QD excited state.

\begin{figure*}
\centering{\includegraphics[width=6in]{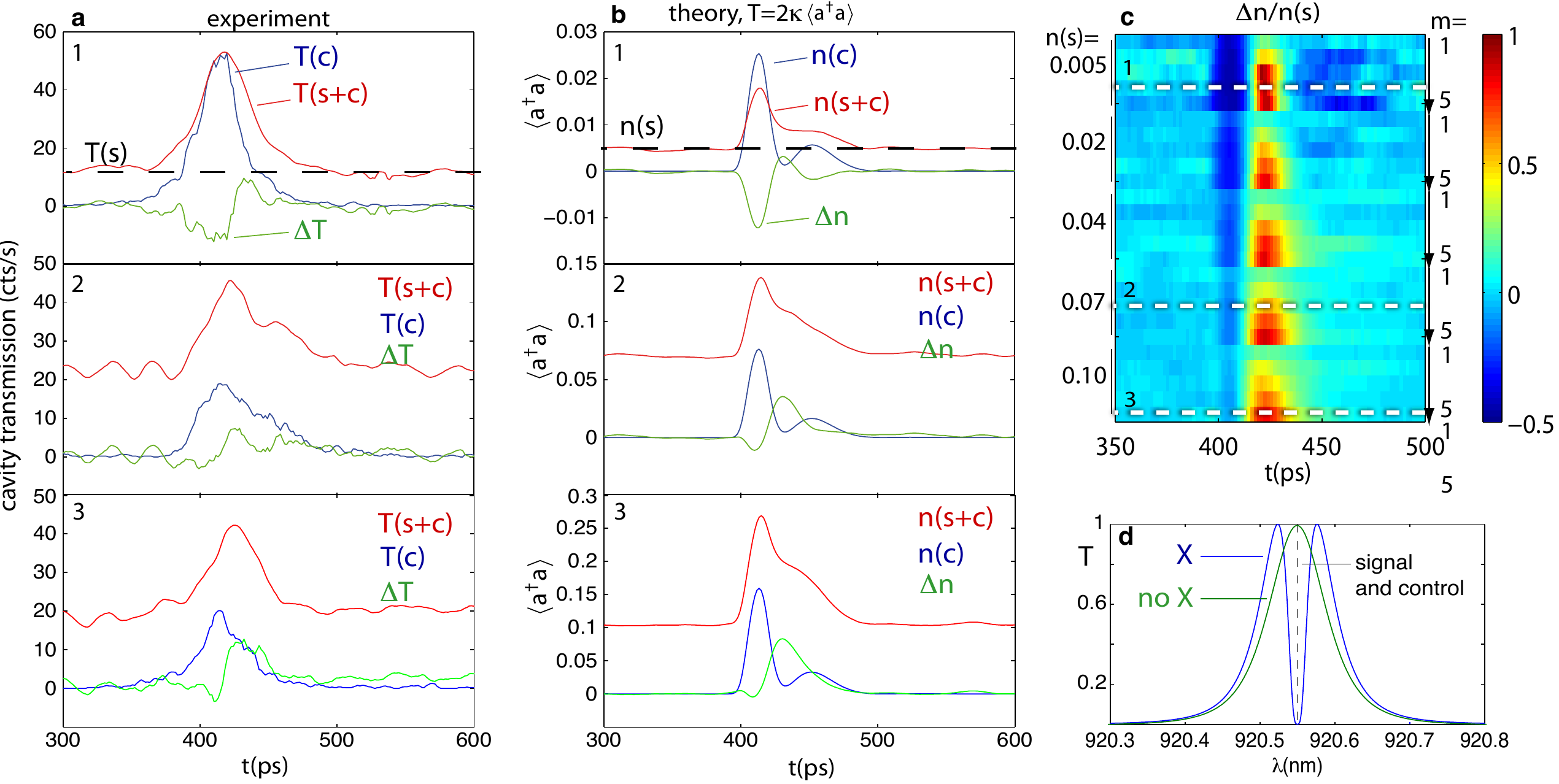}}
\caption{{\footnotesize All-optical switching through the strongly
coupled QD/cavity system. (a) Cavity transmission when the signal
and control are tuned to the cavity, which is resonant with the
QD. As the signal and control input powers are increased, the QD
saturates and results in a net-positive nonlinear transmission.
(b) Plots of the intracavity photon number
$\braket{a^{\dagger}a}$, which gives the transmission by
$T=2\kappa\braket{a^{\dagger}a}$. The calculated intracavity
photon number for the control beam intracavity photon number
$n(c)$, signal and control $n(s+c)$, and the differential photon
number $\Delta n=n(s+c)-n(s)-n(c)$. (c) Theoretical nonlinear
transmission normalized by the signal intensity for a range of
signal and control fields.  The signal intracavity photon number
is equal to $n(s)$ across blocks where the control field is
increased by a multiplier $m$ over the signal field, where
$m=1,2,3,4,5$. The dashed lines correspond to the experimental
conditions (1,2,3) in (a). (d) Calculated transmission with and
without the QD single exciton line. The signal and control field
frequencies are aligned with the QD/cavity resonance. }
}\label{fig:cw_pulse}
\end{figure*}

The coupled QD/cavity system enables a strong interaction between
two weak laser fields. This was previously demonstrated for two
continuous-wave (cw) beams\cite{2008.Science1}.  We now study the
time-resolved dynamics of this interaction between cw `signal' and
pulsed `control' fields. As illustrated in
Fig.\ref{fig:cw_pulse}(d), both fields are within the linewidth of
the cavity resonance. In the experiment, with the QD resonant with
the cavity and the control and signal beams tuned to the bottom of
the transmission dip, we measure the time-resolved transmission of
the control $T(c)$, the signal $T(s)$, and the signal and control
$T(s+c)$ on the streak camera.

We first set the signal and control fields resonant with the tuned
QD-cavity system and attenuate the power so that the average
intracavity photon number is nearly zero; this corresponds to 12
nW for the cw beam and $\sim 0.2$ nW average power for the pulsed
control, both measured before the objective lens. Considering a
coupling efficiency into the cavity of $\eta\sim
3\%$~\cite{2007.Nature1}, this corresponds to an average
intracavity photon number of $\braket{a^{\dagger}a}\sim 0.005$ for
the cw beam and up to 0.025 for the pulsed beam. Here,
$\braket{a^{\dagger}a}$ corresponds to the {\it instantaneous}
cavity photon number. Fig.~\ref{fig:cw_pulse}(a) plots the curves
$T(c), T(s), T(s+c)$, as well as the difference $\Delta
T=T(c+s)-T(c)-T(s)$, which provides a measure of the nonlinear
response of the system. Measurements for two additional sets of
curves are shown in the lower plots in Fig.\ref{fig:cw_pulse}(a).
A stochastic simulation of the quantum master equation (see
Methods) yields the cavity transmission, which is proportional to
the intracavity photon numbers $n(c),n(c+s)$, and $n(s)$, as well
as the nonlinear signal $\Delta n$.  In fitting the data, the
intensities of the pulsed and cw beams were not free parameters,
but were fixed by the experimentally measured optical powers,
using the same coupling efficiency $\eta$. The predicted curves
are plotted in Fig.\ref{fig:cw_pulse}(b) and show good agreement
with the experiment. Fig.\ref{fig:cw_pulse}(c) shows the
theoretical value of the nonlinear coefficients $\Delta n/n(s)$
(normalized by the signal photon number) for the accessible
parameter space given by signal power, control power, and the
cavity parameters $\kappa, \gamma$, and $g$.

\begin{figure}
\centering{\includegraphics[width=.8\linewidth]{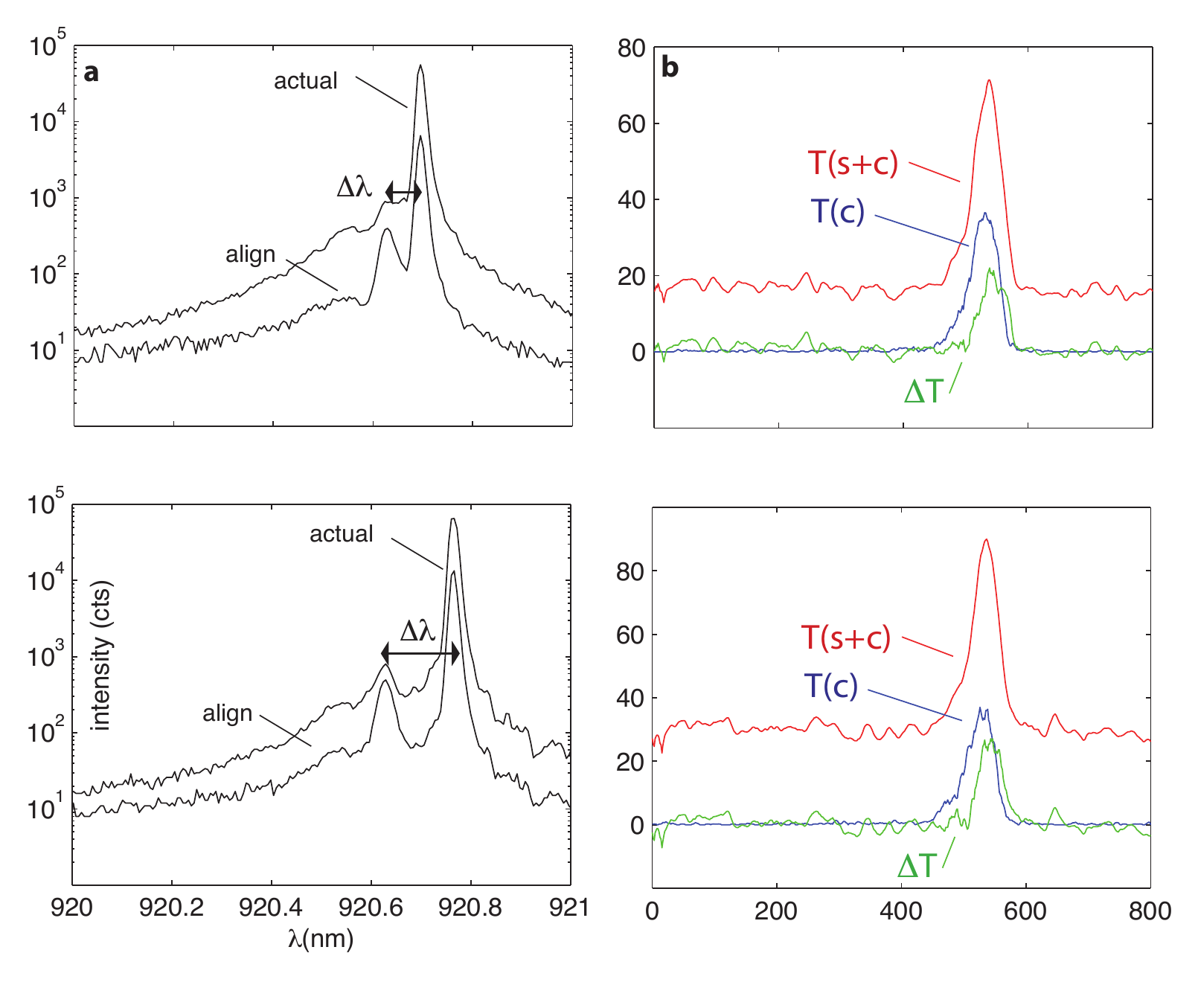}}
\caption{{\footnotesize  Nonlinear interaction between a cw-signal
beam tuned to the cavity and a control pulse detuned by $\Delta
\lambda$. (a) The `align' curve shows the time-averaged control
and the attenuated laser intensities; the `actual' shows the
average powers of the pulsed control and cw signal beams used in
the experiment: $\sim2$ nW for the control, 160 nW for the cw
beam. (b) The control pulse results in a nonlinear transmission of
the signal of $\Delta T$. } }\label{fig:cw_pulse_cross} \vskip2mm
\end{figure}

In applications such as quantum non-demolition measurements, the
interaction between two frequency-detuned probe and signal fields
is of interest.  We therefore repeat our measurements when the
control beam is detuned by up to $\Delta \lambda=0.15$ nm from the
cavity frequency while the cw signal field is maintained on
resonance with the cavity. Two instances are displayed in
Fig.\ref{fig:cw_pulse_cross}, corresponding to detunings of
$\Delta \lambda=\{0.07,0.14\}$ nm for the top and bottom plots,
respectively. After the alignment is done at comparable averaged
cw and pulsed powers (`align' curve in
Fig.\ref{fig:cw_pulse_cross}(a)), the experiment is conducted with
similar intracavity photon number for the cw and pulsed beams.
The experiment shown uses a cw-beam and a pulsed beam with 160 nW
(corresponding to $\braket{a^{\dagger} a}\sim 0.07 $) and $\sim 2$
nW, respectively (measured before the lens).
Fig.~\ref{fig:cw_pulse_cross}(b) displays the transmitted power
acquired on the streak camera, which shows a strong nonlinear
increase for the pulse duration.



\begin{figure}
\centering{\includegraphics[width=3.5in]{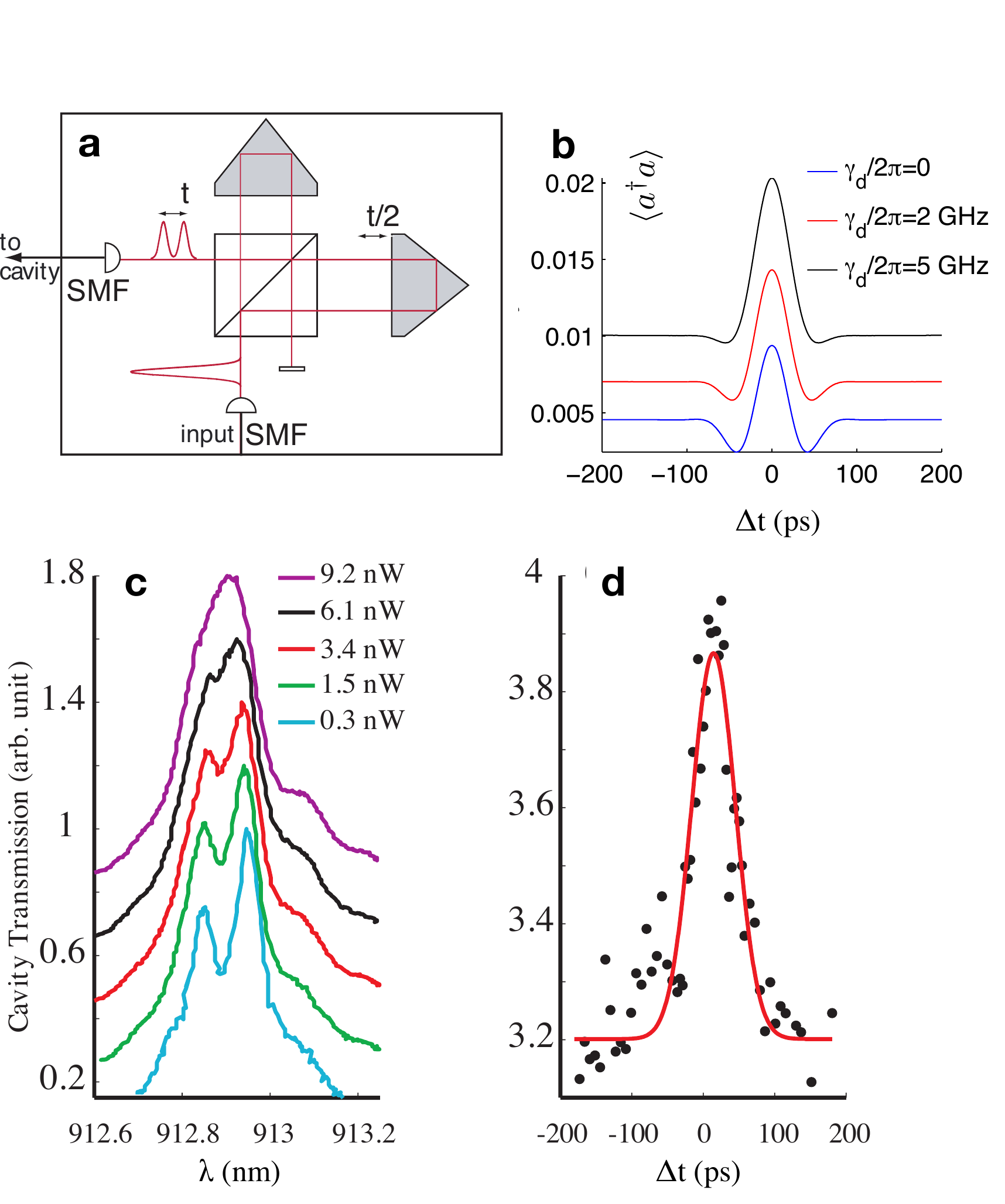}}
\caption{{\footnotesize Interaction of two weak laser pulses through the QD/cavity system. (a) Time-delay setup for producing pulses at a separation of $\Delta t$. (b) Simulated interaction of two laser pulses, represented by the instantaneous intracavity photon number $\braket{a^{\dagger}a}$ as a function of the time delay $\Delta t$ between the two 40 ps long Gaussian pulses. Curves are calculated for a set of different rates of pure QD dephasing\cite{2010.PRL.Englund}, $\gamma_d$, which causes a reduction of the transmission dips before and after the peak. Pure dephasing also causes a blurring of the spectral normal mode splitting\cite{2010.PRL.Englund}, which in turn raises the transmission for increasing $\gamma_d$. (c) Pump-power dependence of the cavity transmission for coincident pulses repeating at 80 MHz.  (d) Signal observed when the cavity-QD system is probed with two 40 ps long pulses as a function of their delay. When the two pulses have a temporal overlap inside the cavity, the QD saturates and the overall cavity reflection increases. The power in the single of the two pulses corresponds roughly to the 3.4nW trace in (c). Best agreement is found with the theoretical plot for a pure dephasing rate $\gamma_d/2\pi \sim 5$ GHz. } }\label{fig:pump_pump} 
\end{figure}

Finally, we consider how two 40-ps laser pulses, resonant with the
cavity and having a relative delay of $\Delta t$ interact through
the cavity. The pulse pair is generated using the delay setup of
Fig.\ref{fig:pump_pump}(a). To average out interference between
the two pulses, we detune them by 40 MHz and average over many
pulse pairs (this detuning is very small compared to the pulse
bandwidth). Numerical integration of the Master equation predicts
an increased reflection when both pulses are simultaneously
coupled to the cavity; this is shown for a particular choice of
power in Fig.\ref{fig:pump_pump}(b). This experiment is performed
on a different QD-PC cavity system with similar parameters
$\{\kappa,g,\gamma\}/2\pi=\{27.2,21.2,0.1\}$ GHz; the temperature
is 38K. Fig.\ref{fig:pump_pump}(c) plots the time-averaged
reflected signal observed on a spectrometer for coincident pump
pulses with average power of both pulses increasing from 0.3 nW to
9.2 nW before the objective lens. It is evident that for powers
beyond $\sim 9$ nW, the polariton mode splitting disappears as the
QD is saturated. This suggests that the QD-cavity system acts as a
highly nonlinear system that increases its transmission for
coincident pulses. This is what we find in
Fig.\ref{fig:pump_pump}(d): the cavity transmission rises by 22\%
when the pulses are coincident at $\Delta t=0$. This transmission
peak agrees with the theoretical prediction of a $\sim 40$ ps
duration, as shown in the red curve.  In the theory plot, we also
observe reduced transmission at a non-zero time delay. We find
that by including a pure QD dephasing term into the master
equation model, the transmission dips are diminished, as observed
in the experimental results. Pure QD dephasing was previously
shown to play an important role in the QD-cavity system, including
off-resonant interaction between the exciton and cavity
mode\cite{2010.PRL.Englund}. The best fit to experimental data is
found for a dephasing rate of $\gamma_d/2\pi\approx 5$ GHz.

In conclusion, we have described the interaction between
time-varying control and signal fields via a strongly coupled
QD/cavity system. A strong nonlinear response exists even at low
intensity corresponding to mean intracavity photon numbers below
one. This all-optical interaction is promising for quantum
information processing with optical
nonlinearities\cite{1995.PRL.Turchette-Kimble,1995.PRA.Chuang-Yamamoto.simple_QC,2005.PRA.Munro.QND_high_efficiency_abbr}.
The large nonlinearity may also be of use in classical all-optical
signal processing\cite{2009.PRA.Mabuchi.cavity_QED_switch} -- for
example, for the implementation of all-optical logic gates
operating at the single- or few-photon level. The QD-cavity system
is ideal for on-chip integration and can easily operate with
repetition rates up to a tens of GHz.

Financial support was provided by the Office of Naval Research
(PECASE Award), National Science Foundation, and Army Research
Office. A.M. was supported by the SGF (Texas Instruments Fellow).
Work was performed in part at the Stanford Nanofabrication
Facility of NNIN supported by the National Science Foundation.
D.E. acknowledges support by the AFOSR YIP and the Sloan Research
Fellowship.

\appendix
\section{Analytical Model} The QD is described as a two level
system with a ground state $\ket{g}$ and an excited state
$\ket{e}$. The system is characterized by a dipole decay rate
$\gamma$, a cavity field decay rate $\kappa$, and a QD-cavity
field coupling at the Rabi frequency $g$. The driving field is
described by $\Omega(t)$. The interaction of the laser field with
the coupled QD-cavity system is described by the Jaynes Cummings
Hamiltonian (in a frame rotating at the laser frequency)
\begin{equation}
H=\Delta\omega_c a^\dag a +\Delta\omega_d \sigma^\dag
\sigma+ig(a^\dag \sigma
-a\sigma^\dag)+i\sqrt{\kappa}\Omega(t)(a-a^\dag)
\end{equation}
where $\Delta\omega_c = \omega_c-\omega_l$ and $\Delta\omega_d =
\omega_d-\omega_l$ are respectively the cavity and dot detuning
from the driving laser.

To incorporate the incoherent losses in the system we find the
Master equation, given by
\begin{equation}
\label{Maseq} \frac{d\rho}{dt}=-i[H,\rho]+ 2\kappa
\mathcal{L}[a]+2\gamma
\mathcal{L}[\sigma]+2\gamma_d\mathcal{L}[\sigma^\dag\sigma]
\end{equation}
where $\rho$ is the density matrix of the coupled QD/cavity system
and $\gamma_d$ is the quantum dot pure dephasing rate.
$\mathcal{L}[D]$ is the Lindblad operator corresponding to a
collapse operator $D$. This is used to model the incoherent decays
and is given by:
\begin{equation}
\mathcal{L}[D]= D\rho D^\dag-\frac{1}{2}D^\dag D
\rho-\frac{1}{2}\rho D^\dag D
\end{equation}
The Master equation is solved using Monte Carlo integration
routines (for the mixed CW and pulsed case) and by using the
numerical integration routines (for the two pulse switching case)
provided in the quantum optics toolbox\cite{TanMATLAB}. For the
two pulse switching, we assume a Gaussian pulse-shape for the
pulses.

\end{document}